\documentstyle[12pt,aps]{revtex}
\tighten
\begin{document}
\title{{\Large {\bf 
Chiral NN interactions in nuclear matter\bigskip}}}
\author{Boris Krippa\bigskip}
\address{ Department of Physics and Astronomy, Free University
of Amsterdam,\\ De Boelelaan 1081, 1081 HV Amsterdam.\\}
\maketitle
\vspace{1cm}
\begin{abstract}
 We consider an effective field theory of NN system
in nuclear medium. The shallow bound states, which complicate the 
effective field theory analysis and lead to the large scattering length
in the vacuum case
do not exist in matter. We show that  
the next-to-leading order terms in the chiral expansion of the 
effective NN potential can  be interpreted as corrections so that
the expansion is systematic. It is pointed out however
that it is still useful to treat the problem nonperturbatively since
it may allow for the consideration of the nuclear systems with the density
smaller that the normal nuclear matter one.
The potential energy per particle is
calculated. The possible directions in 
constructing the chiral theory of nuclear matter are outlined.

\end{abstract}

\vskip0.9cm


Effective Field Theory (EFT)
has become a popular tool for studying nuclear interactions.
EFT is based on the idea to use the Lagrangian
 with the appropriate effective degrees of freedom instead
of the fundamental ones in the low-energy region (for review of EFT see,
for example \cite {Ma95}). This Lagrangian should include all possible terms 
allowed by the symmetries of the underlying QCD.  The states which
 can  be treated as heavy, compared to the typical energy scale involved,
are integrated out. They are hidden in the Low Energy Effective
Constants (LEC's) of the corresponding Lagrangian.
 The physical amplitudes can be represented as the sum 
of certain graphs, each of them being of a given order in $Q/\Lambda$,
where $Q$ is a typical momentum scale and $\Lambda$ is a parameter 
reflecting the scale of the short range physics.
 The relative contribution of each graph can  roughly be estimated
using chiral counting rules \cite {We79}. The relevant degrees of freedom in
the nuclear domain are nucleons and pions. In the case
 of meson-meson \cite {GL} and meson-nucleon\cite {Ec}
interactions the perturbative chiral expansion can be organized in a 
consistent way. However, being applied to the NN system EFT encounters
serious problem which is due to existence of the  bound
states near threshold \cite {We91}. It results in the large nucleon-nucleon
scattering length and makes the perturbative expansion divergent. Weinberg 
suggested \cite {We91} to apply chiral counting rules to the certain class 
of the irreducible diagrams which should  then be summed up to infinite 
order by solving the Lippmann-Schwinger (LS) equation. 
The irreducible diagrams can be treated as the effective potential
 in this case. Different aspects of the chiral NN problem 
have been discussed since then \cite {Ka}. The concept of EFT
has also intensively been used to study  nuclear matter 
\cite {Se97,Ly,Lu,Fr}. In \cite {Se97} the effective
 chiral Lagrangian was constructed and the ``naturalness'' of the
 effective coupling constants has been demonstrated. The possible
counting rules for nuclear matter have been discussed in \cite {Lu}.
These two lines of development of the chiral nuclear physics are in some sense
similar to the tendencies existed some time ago in  conventional nuclear
physics with the phenomenological two body forces. On the one hand,
 the phenomenological NN potentials were used to describe
 nucleon-nucleon cross sections
and phase shifts. On the other hand,  nuclear mean field approaches
 provided a reasonable description of the bulk
 properties of nuclear matter.
 The unification of these two approaches then led to the famous
Bethe-Goldstone (BG) equation \cite {Be} for the G-matrix which is an 
analog of  scattering T-matrix, satisfying the LS equation.  
It is therefore reasonable to follow the same strategy and, being
 equipped with
the chiral theory of NN interaction in vacuum, try to construct the chiral
G-matrix, describing the effective interactions of two nucleons in medium.
One can easily see the qualitative difference between vacuum and medium cases.
 In nuclear medium because of Pauli blocking  the intermediate 
states with the momenta
less than Fermi momentum $p_F$ are forbidden. Therefore, the nucleon propagator
does not exhibit a pole. Moreover, the shallow bound or virtual NN states,
which constitute the main difficulty of the problem in vacuum, simply do not
exist in nuclear matter because of  interaction of the NN pair with
 nuclear mean
 field. It means that the effective scattering length becomes considerably
smaller compared to the vacuum one. 
 The value of a scattering length is determined
by the position of the singularity, nearest to the physical region.
In the vacuum case, for example, the virtual deuteron bound state is very close
to the NN threshold leading to the unnaturally large scattering
length. The moderate value of the in-medium scattering length would
 indicate, in some sense, that the typical scale 
 of the NN interactions 
gets ``more natural'' in nuclear matter.\\        
We start from the standard nucleon-nucleon effective chiral Lagrangian
which can be written as follows
\begin{equation}
{\cal L}=N^\dagger i \partial_t N - N^\dagger \frac{\nabla^2}{2 M} N
- \frac{1}{2} C_0 (N^\dagger N)^2\\ 
-\frac{1}{2} C_2 (N^\dagger \nabla^2 N) (N^\dagger N) + h.c. + \ldots.
\label{eq:lag}
\end{equation}
We consider the simplest case of the NN scattering in the $^{1}S_0$ state
 and assume
zero total 3-momentum of NN pair in the medium. The inclusion of
the nonzero total 3-momentum does not really change anything 
qualitatively and only makes the calculations technically more involved.   
The G-matrix is given by
\begin{equation}
G(p',p)=V(p',p) + M \int \frac{dq q^2}{2 \pi^2} \, V(p',q) 
\frac{\theta(q-p_F)}{M(\epsilon_{1}(p) +\epsilon_{2}(p')) - q^2} G(q,p),
\label{eq:LSE2}
\end{equation}
Here $\epsilon_1$ and $\epsilon_2$ are the single-particle 
energies of the bound nucleons.
They  are affected by the nuclear mean field. In nuclear medium such corrections lead to 
the nucleon  effective mass slightly different from that in free space.
We used the value $M = 0.8 M_0$, where $M_0$ is the nucleon mass in vacuum.
One notes that this value is close to one usually accepted in nuclear 
mean field theories. 
 The standard strategy of  treating the chiral NN problem in vacuum
is the following. One computes amplitudes up to a given chiral order
in the terms of the effective constants $C_0$ and $C_2$ which are then 
determined by comparing the calculated amplitude with some experimental data.
 Having these constants fixed one
can calculate the other observables. 
We will follow the similar strategy  in the nuclear matter case  
and proceed as follows.
We choose exactly solvable separable potential with parameters adjusted
to the value of the potential energy per particle in nuclear matter.
 Then we  solve
 the BG equation
with the effective constants $C_0$ and $C_2$. The numerical values 
of these constants are determined comparing the phenomenological
and EFT G-matrix at some fixed kinematical points.
The check of consistency we used is the 
 difference between $C_0$'s
determined in the leading and subleading orders. 
 If the difference between the values of $C_0$ needed to fit the data  
in leading and subleading order is of higher  order then the procedure 
of truncation of the standard chiral expansion is justified. 
In the vacuum case
the corresponding difference was found to be large \cite{Co97}.
Using a simple separable potential
\begin{equation}
V=-\lambda |\eta\rangle\langle\eta|
\end{equation}
with the form factors

\begin{equation}
\eta(p)=\frac{1}{(p^2 + \beta^2)^{1/2}}
\end{equation}
 One can easily get 
\begin{equation}
\frac{1}{T(k,k)}=
{V(k,k)^{-1}}\left[1 - M_0 \int \frac{dqq^2}{4 \pi^{2}} \, \frac{V(q,q)}
{{k^2}- q^2}\right] 
\label{eq:sep}
\end{equation}
The experimental values of scattering length $a$ and effective radius $r_e$
are
\begin{equation}
a=-23.71\pm 0.013\, {\rm fm}\qquad {r_e}=2.73\pm 0.03\, {\rm fm}.
\label{eq:exp}
\end{equation}
These values can be reproduced if we choose
\begin{equation}
\lambda = 1.95\,\qquad \beta = 0.8\, {\rm fm}
\end{equation}
The solution of the BG equation for the separable potential is a simple
generalization of the one for the LS equation 
\begin{equation}
G(k,k)=-
\eta^{2}(k)\left[\lambda^{-1}+ 
\frac{M}{2 \pi^{2}} \int{dqq^2} \, \frac{\theta(q-p_F) \eta^{2}(q)}{{k^2}- q^2}\right]^{-1}
\end{equation}
However, the phenomenological G-matrix with the parameters determined
from the effective range expansion fit leads to the  somewhat lower
the potential energy  than the usually accepted value $\sim$ -16 MeV.

To get a better fit we choose 
\begin{equation}
\lambda = 2.4\,\qquad \beta = 1.1\, {\rm fm}
\end{equation}
These values are fairly close to the vacuum ones and provide 
the potential energy per particle in a good agreement with the empirical value. The parameters
 $ \lambda$ and $\beta$ being substituted in the G-matrix lead  to  
$a_m\simeq r_m\simeq 0(1)$, where $a_m$ and
  $r_m$ are the in-medium analogs of scattering length and effective radius.
 One notes that effective radius is much less affected by the
medium effects. It is quite natural since the value of the effective radius is only
weakly sensitive to the bound state at threshold and  is of the ``almost natural'' size
already in the vacuum case. 
 The absolute value of the  in-medium scattering length is considerably reduced 
compared to the vacuum one. It clearly indicates that, as expected, the 
shallow virtual nucleon-nucleon bound state is no longer present in nuclear medium.
Thus,  one can avoid significant part of the difficulties typical for the chiral NN
problem in vacuum. Having determined the phenomenological G-matrix
one can now solve the BG equation using leading and sub-leading
orders of the NN effective chiral Lagrangian. The solution is similar to 
the vacuum case \cite{Co97} and can be represented as follows  

\begin{equation}
\frac{1}{G(k,k)}=\frac{(C_2 I_3(k,p_F) -1)^2}{C_0 + C_2^2 I_5(k,p_F)
 + {k^2} C_2 (2 - C_2 I_3(k,p_F))} - I(k,p_F),
\label{eq:Tonexp}
\end{equation}
where we defined

\begin{equation}
I_n \equiv -\frac{M}{(2 \pi)^2} \int dq q^{n-1}\theta(q-p_F).
\label{In}
\end{equation}
 
and  

\begin{equation}
I(k) \equiv \frac{M}{2 \pi^{2}}\int dq  \, \frac{q^2\theta(q-p_F)}
{{k^2}- {q^2}}.\label{eq:IEdef}
\end{equation}
These integrals are divergent so the renormilization should be carried out.
The procedure used is similar to that adopted in Ref. \cite {Ge} to study the 
EFT approach to the NN interaction in vacuum. We subtract the divergent integrals at
some kinematical point $p^2 = -\mu^2$. After subtraction the renormalized G-matrix takes the form

\begin{equation}
\frac{1}{G^{r}(k,k)}=\frac{1}{C^{r}_{0}(\mu) 
 +  2{k^2} C^{r}_{2}(\mu)} +\frac{M}{4\pi}[p\log\frac{p_F - p}{p_F + p}
 - i\mu\log\frac{p_F - i\mu}{p_F + i\mu}] ,
\label{eq:GR}
\end{equation}
One notes that in the $p_F \rightarrow 0$ limit  the vacuum chiral NN amplitude is recovered.
We choose the value $\mu$ = 0 as a subtraction point. The $\mu$ dependence of LEC's is 
governed by the renormalization group equation. Now one can determine the LEC's by equating
the EFT and phenomenological G-matrices at some kinematical points. We used the values 
$p = \frac{p_F}{2} ; \frac{p_F}{3}$ as such points. The assumed value of the Fermi-momentum
is $p_F$ = 1.37 fm. In the following we will omit the label ``r'' implying that we always deal with 
renormalized quantities.  
We found $C_0 = -1.86 fm^2$ in LO. In NLO one gets $C_0= 2.64 fm^2$ and $C_2 = 0.84 fm^4$
so that the inclusion of the NLO corrections  give rise to the approximately 40$\%$ change
in the value of  $C_0$. It indicates that the chiral expansion is systematic in a sense that adding of the NLO 
terms in the effective Lagrangian results in a ``NLO change'' of the coefficients which have already 
been determined at LO. The natural size of the in-medium scattering length and moderate changes
experienced by the coupling constant $C_0$ might, in principle, indicate the possibility
of the perturbative calculations.   However, in spite of
this, it is still more
useful to treat this problem in the nonperturbative manner. There are few 
reasons for the nonperturbative treatment. Firstly, the corrections themselves
are quite significant. Secondly, the overall (although distant)
goal of the EFT description is to derive both nuclear matter and the 
vacuum NN amplitude from the same Lagrangian, However, it is hard to say 
at what densities the dynamics becomes intrinsically nonperturbative, so it is 
better to treat the problem nonperturbatively from the beginning.
The nonperturbative treatment may also turn out important to get the 
correct saturation curve since at some density lower than the normal
 nuclear one the scattering length starts departing from its natural value
and some sort of the nonperturbative approach becomes inevitable.
 Thirdly, in the processes
involving both the nonzero density and temperature, such as heavy ion
collisions, the value of the Fermi-momentum can effectively be lowered again
making the nonperturbative treatment preferable. 

Let's now calculate the potential energy per particle using the expression 
for the in-medium chiral NN scattering amplitude. The potential energy
of nuclear matter can be evaluated from
 \begin{equation}
U_{tot} = \frac{1}{2}\sum_{\mu,\nu}<\mu\nu|G(\epsilon_\mu + \epsilon_\nu)
|\mu\nu-\nu\mu)
\end{equation}
The summation goes over the states with momenta below $p_F$.
Here it is seen that $G$ amplitude plays the role of an effective chiral two-body 
interaction in nuclear medium. The calculations using the lowest 
order $G$-matrix result in the value $\frac{U(^1S_0)}{A}\simeq -17 MeV$.
The inclusion of the next-to-leading order corrections gives rise to the 
value $\frac{U(^1S_0)}{A}\simeq -13.1 MeV$.
One notes that both $\frac{U(^1S_0)}{A}$ and $C_0$ experience corrections of the
same order when NLO terms are included in the effective Lagrangian. The 
similar calculations done in the triplet s-wave channel give rise to to the value
$\frac{U(^3S_1)}{A}\simeq -17.3 (-13.2)$ MeV in LO (NLO). The values of the potential
 energy obtained with 
chiral approach looks quite reasonable although they are somewhat smaller 
than the standard values usually obtained in the calculations with the 
phenomenological two-body forces \cite {Tab}. One can therefore conclude that
there is still a room for both pionic effects and three particle correlations which
should be included in a chirally invariant manner.

The validity of the EFT description is restricted by some cutoff parameter 
reflecting the short range physics effects. Its value deserves some comments 
in the context of applying of the  EFT methods to nuclear matter. The scale 
where the EFT treatment ceases to be valid should approximately correspond 
to the scale of the short range correlations (SRC),
 that is, $\sim 2.5 fm^{-1}$.
The description of SRC is hardly possible in the framework of EFT so the value 
$\Lambda \sim 2 fm^{-1}$ might put natural constraint on the EFT description 
of nuclear matter. To make the chiral expansion meaningful the chiral counting rules 
in nuclear matter must be established. This is still open problem. However, 
the above obtained results suggest that the relevant expansion parameter 
could be something like $\frac{<p>}{\Lambda} \sim \frac{<m_\pi>}{\Lambda} \sim 0.3 -0.4$.
 Of course, until pion effects are taken into account this estimate can only be suggestive.
Moreover, many other things remain to be done to make the qualitative description of nuclear matter possible. Beside pionic effects one needs to include many body forces and formulate
chiral counting rules. One should also find a way to remove off-shell ambiguities
order by order and calculate the nucleon self-energy up to a given chiral order
to make EFT description of nuclear matter fully consistent.

\section*{Acknowledgments}

Author is very grateful for the support
and warm hospitality from SRCSSM at the University of Adelaide
 where the initial part of this work was done.

\end{document}